\documentclass[%
 reprint,
 amsmath,amssymb,
 aps,
 pra,
]{revtex4-1}

\usepackage{graphicx}
\usepackage{dcolumn}
\usepackage{bm}
\usepackage{float}
\usepackage{xcolor}
\usepackage{amsmath}
\usepackage{mathtools}
\usepackage{etoolbox}
\usepackage{physics}
\apptocmd{\thebibliography}{\raggedright}{}{}
\usepackage{amsmath}
\usepackage[titletoc]{appendix}

\usepackage{soul}

\def \beq {\begin{equation}}
\def \eeq {\end{equation}}

\begin{document}

\title {Coalescence of non-Markovian dissipation, quantum Zeno effect and non-Hermitian physics, in a simple realistic quantum system}

\author{G. Mouloudakis$^{1,2}$}
 \email{gmouloudakis@physics.uoc.gr}

\author{P. Lambropoulos$^{1,2}$}%

\affiliation{${^1}$Department of Physics, University of Crete, P.O. Box 2208, GR-71003 Heraklion, Crete, Greece
\\
${^2}$Institute of Electronic Structure and Laser, FORTH, P.O.Box 1527, GR-71110 Heraklion, Greece}

\date{\today}

\begin{abstract}

Diagonalization of the effective Hamiltonian describing an open quantum system is the usual method of tracking its exceptional points. Although, such a method is successful for tracking EPs in Markovian systems, it may be problematic in non-Markovian systems where a closed expression of the effective Hamiltonian describing the open system may not exist. In this work we provide an alternative method of tracking EPs in open quantum systems, using an experimentally measurable quantity, namely the effective decay rate of a qubit. The quantum system under consideration consists of two non-identical interacting qubits, one of which is coupled to an external environment. We develop a theoretical framework in terms of the time-dependent Schrödinger equation of motion, which provides analytical closed form solutions of the Laplace transforms of the qubit amplitudes for an arbitrary spectral density of the boundary reservoir. The link between the peaked structure of the effective decay rate of the qubit that interacts indirectly with the environment, and the onset of the Quantum Zeno Effect, is discussed in great detail revealing the connections between the latter and the presence of exceptional points. Our treatment and results have in addition revealed an intricate interplay between non-Markovian dynamics, Quantum Zeno effect and non-Hermitian physics.

\end{abstract}

\maketitle

\section{Introduction}

The dissipative dynamics of open quantum systems coupled to non-Markovian reservoirs is a multifaceted field of fundamental, as well as practical importance. It pertains to a broad class of problems, ranging from quantum information processing to non-equilibrium statistical mechanics. The effective Hamiltonian describing an open quantum system is by necessity non-Hermitian, which brings up its possible connection with non-Hermitian physics, exceptional points and related questions, in that field of wide-ranging interest and activity. In both of those fields and from different angles, the quantum Zeno effect (QZE) has been found to be a major participant. Although the latter entered quantum physics as a curiosity, it has been found to play an uncanny role in the protection against dissipation. The number of past and current articles in each of the above three fields is immense. Yet, the synergy of phenomena related to those fields does not seem to have been noticed, let alone explored. Our recent work \cite{ref1} on quantum dissipation in non-Markovian environments has steered us to a type of problem in which that synergy has been found to be astonishingly revelatory. The treatment of that problem and its consequences is the purpose of the present article. Before embarking on the discussion of formulation, computation and results, we need to provide a brief outline of the background and past activity, in each of the above three fields.

Dissipation is essentially inevitable in any process involving a quantum system, arising from its interaction with the environment, referred to as reservoir, or even a class thereof. A reservoir is characterized by a specific spectral density. Depending on whether that spectral density is smooth or exhibits a peaked behaviour, at least in a range of energies encompassing the energy of the system, the reservoir is usually referred as Markovian or non-Markovian, respectively. Physically speaking, the term Markovian refers to reservoirs for which the Markov approximation is valid. This implies that any excitation transferred from the system to the reservoir is irreversible i.e. practically lost forever \cite{ref2}. On the other hand, for non-Markovian reservoirs, although eventual loss is also present, the excitation may be transferred back to the system. This exchange of excitation between system and reservoir lasts for finite times, whose length depends on the spectral density of the latter. The length of that time does in fact characterize the so-called Markovianity of the particular reservoir \cite{ref3,ref4}.

Although the interaction of a quantum system with an external environment does ultimately lead to dissipation, there is an important effect which, depending on the relative parameters of the compound system, may lead to protection against such types of dissipation. That effect known as the "Quantum Zeno Effect" (QZE), reflects the possibility of the environment to freeze the dynamics of the quantum system (or part of it) through frequent projective measurements of its state \cite{ref5}. The QZE, the regions of its onset and its implications have been studied in many different contexts such as in circuit-QED systems \cite{ref6}, in ultra-cold atoms \cite{ref7} and in 1D “hybrid” quantum circuit models \cite{ref8}, while many experiments have confirmed the possibility of freezing the evolution of the quantum state via such a mechanism \cite{ref9,ref10,ref11,ref12,ref13,ref14,ref15}.

A number of important studies have also pointed out the potential role of the QZE in the protection of quantum information between correlated qubits \cite{ref16,ref17,ref18,ref19}. The results suggest that repeated projective measurements on a system of entangled qubits can lead to the preservation of entanglement, independently of the state in which the system is initially encoded. This effect appears when the state of the system evolves in a multidimensional sub-space, usually referred to as the Zeno subspace \cite{ref20,ref21}. Although fast repeated projective measurements directly on the system may freeze its evolution, this method may be somewhat restrictive for the implementation of quantum information processing tasks, where additional operations on the system may be necessary. An alternative approach relies on "indirect" measurements, where the apparatus does not act directly on the system, but detects a signal mediated by some field with which it interacts \cite{ref22}. That work has, however, given rise to serious reservations as to the possibility of the occurrence of the QZE in such configurations \cite{ref23,ref24,ref25,ref26,ref27}. On the other hand, it has been demonstrated that the QZE does not necessarily require projective measurements, as it may also be induced through continuous strong couplings \cite{ref15,ref28,ref29,ref30,ref31}.

In recent work \cite{ref32}, W. Wu and H.-Q. Lin have investigated the QZE in dissipative systems beyond the Markov, the rotating-wave and the perturbative approximations, in the context of a spin-boson model which describes the interaction between a spin system and a bosonic bath. Their study suggested that the non-Markovian character of the bath may be favorable for the accessibility of the QZE in such systems, as it may prolong the quantum Zeno time and lead to multiple Zeno-anti-Zeno crossover phenomena.

At the same time, the transitions to the Quantum Zeno regime have been recently shown to be linked with the $\mathcal{P} \mathcal{T}$ symmetry breaking of the non-Hermitian Hamiltonian which describes the open quantum system \cite{ref33,ref34,ref35,ref36,ref37}. The boundary between the unbroken and broken $\mathcal{P} \mathcal{T}$ symmetry of a Hamiltonian describing an open quantum system \cite{ref38} is marked by the presence of exceptional points (EPs) \cite{ref39,ref40} where two or more eigenvalues coalesce, while their corresponding eigenvectors become parallel. It has also been demonstrated that the onset of the QZE is marked by a cascade of transitions in the system dynamics, as the strength of a continuous partial measurement on the open system is increased \cite{ref41}.

Tracking of EPs in open quantum systems is of crucial importance,  since the system appears to exhibit enhanced sensitivity in their vicinity \cite{ref42,ref43,ref44}. For $N^{th}$ order EPs, i.e. EPs that mark the coalescence of $N$ eigenvalues, the sensitivity in the response of the system to small perturbations in parameter space has been confirmed to become more pronounced as $N$ is increased \cite{ref44,ref45,ref46}.

Although in open Markovian systems, tracking EPs through diagonalization of the corresponding effective Hamiltonian is a rather easy theoretical task, that method is rather problematic in non-Markovian systems, for which it may not even be possible to construct an effective Hamiltonian describing the open system. In that case, alternative methods capable of tracking EPs indirectly, without  the need of finding the eigenvalues of the open system, should be sought.

In this work we have developed such a method, illustrating its advantage in a simple open quantum system consisting of two interacting qubits, one of which is coupled to an external environment. Our formulation allows for the derivation of closed form analytical expressions of the Laplace transforms of the qubit amplitudes, for arbitrary spectral densities of the boundary reservoir, enabling the study of the effects of various types of reservoir spectral densities. If the qubit not directly coupled to the environment is initially in its excited state, then as the coupling between the remaining qubit and the environment increases, we observe a phase transition to the Zeno regime, resulting to increased protection against the population dissipation that the environment inevitably induces. A glimpse of this effect was reported recently in an recent paper of ours, for a system of $XX$ spin chains boundary driven by non-Markovian environments, where the total population of the chain was found to become increasingly protected against dissipation, for sufficiently large boundary couplings \cite{ref1}. Here, we investigate the connection between these types of phase transitions and the presence of exceptional points both for Markovian and non-Markovian environments, using an experimentally measurable quantity, namely the effective decay rate of the qubit that does not communicate directly with the bath. Based on a comparative analysis with the case of a Markovian reservoir, for which the system is diagonalizable, we argue that the effective decay rate may be used as a method for tracking the onset of the QZE in an non-Markovian open quantum system, as well as its EPs.

The rest of the paper is organized as follows: In section II we outline the theoretical formulation of the problem, in the case of two non-identical interacting qubits, one of which is coupled to an external environment characterized by an arbitrary spectral density. In section III we provide the results of our study as well as a discussion related to the effects associated with the onset of QZE in the population dynamics of the qubits and its link to exceptional points, while in section IV we summarize the results of our work, providing also some concluding remarks. 

\section{Theory}

Our system consists of two non-identical qubits and an environment characterized by a specific spectral density $J \left( \omega \right)$. The two qubits are interacting with a coupling strength $\mathcal{J}$ while the environment is interacting with the second qubit with a coupling strength $g$. For the sake of simplicity we assume that the coupling strengths $\mathcal{J}$ and $g$ are real numbers. A schematic representation of our system is depicted in Fig. 1.

The Hamiltonian of our system $\hat{\mathcal{H}}=\hat{\mathcal{H}}_S+\hat{\mathcal{H}}_E+\hat{\mathcal{H}}_{I}$ consists of three parts; namely, the Hamiltonian $\hat{\mathcal{H}}_S$ which describes our system of qubits and their mutual interaction, the Hamiltonian of the bosonic environment $\hat{\mathcal{H}}_E$ and the interaction Hamiltonian $\hat{\mathcal{H}}_{I}$ which describes the interaction between the second qubit and the environment. These three Hamiltonian terms are given by the expressions ($\hbar=1$):

\begin{subequations}
\beq
\begin{split}
\hat{\mathcal{H}}_S   = & \omega_g \ket{g}_1 \prescript{}{1}{\bra{g}} + \omega_e \ket{e}_1 \prescript{}{1}{\bra{e}} + \omega_g' \ket{g}_2 \prescript{}{2}{\bra{g}} + \\
& \omega_e' \ket{e}_2 \prescript{}{2}{\bra{e}}  + {\mathcal{J}} \left( \hat{\sigma}_1^{+} \hat{\sigma}_{2}^{-} + \hat{\sigma}_1^{-} \hat{\sigma}_{2}^{+} \right) ,
\end{split}
\label{H_S}
\eeq

\beq
\hat{\mathcal{H}}_E =  \sum_\lambda \omega_{\lambda} \hat{a}_{\lambda}^{{E}{\dagger}} {\hat{a}}_{\lambda}^{E} ,
\eeq

\beq
\begin{split}
\hat{\mathcal{H}}_I = \sum_\lambda g \left( \omega_{\lambda} \right) \left( \hat{a}_{\lambda}^{E} \hat{\sigma}_2^{+} +\hat{a}_{\lambda}^{{E}{\dagger}} \hat{\sigma}_2^{-} \right),
\end{split}
\eeq
\end{subequations}
where $\omega_g$ and $\omega_e$ are the energies of the ground and excited state of first qubit, respectively, $\omega_g'$ and $\omega_e'$ are the energies of the ground and excited state of second qubit, respectively, $\omega_{\lambda}$ is the energy of the $\lambda^{th}$ mode of the environment, $\hat{\sigma}_j^{+}= \ket{e}_j \prescript{}{j}{\bra{g}}$ and $\hat{\sigma}_j^{-}= \ket{g}_j \prescript{}{j}{\bra{e}}$, $j=1, 2$, are the qubit raising and lowering operators, respectively, while $\hat{a}_{\lambda}^{E}$ and $\hat{a}_{\lambda}^{{E}{\dagger}}$ are the quantum annihilation and creation operators of the environment.

The wavefunction of the whole system in the single-excitation space can be expressed as

\beq
\ket{\Psi (t)}   = c_1(t) \ket{\psi_1}  + c_2 (t) \ket{\psi_2} + \sum_{\lambda} c_{\lambda}^{E}(t) \ket{\psi_{\lambda}^{E}},
\eeq
where,

\begin{subequations}
\beq
\ket{\psi_1} = \ket{e}_1 \ket{g}_2 \ket{0}_{E},
\eeq

\beq
\ket{\psi_2} = \ket{g}_1 \ket{e}_2 \ket{0}_{E},
\eeq

\beq
\ket{\psi_{\lambda}^{E}} = \ket{g}_1 \ket{g}_2 \ket{0 0 \dots 0 1_{\lambda} 0 \dots 0 0}_{E}.
\eeq 

\end{subequations}

\begin{figure}[t] 
	\centering
	\includegraphics[width=8.65cm]{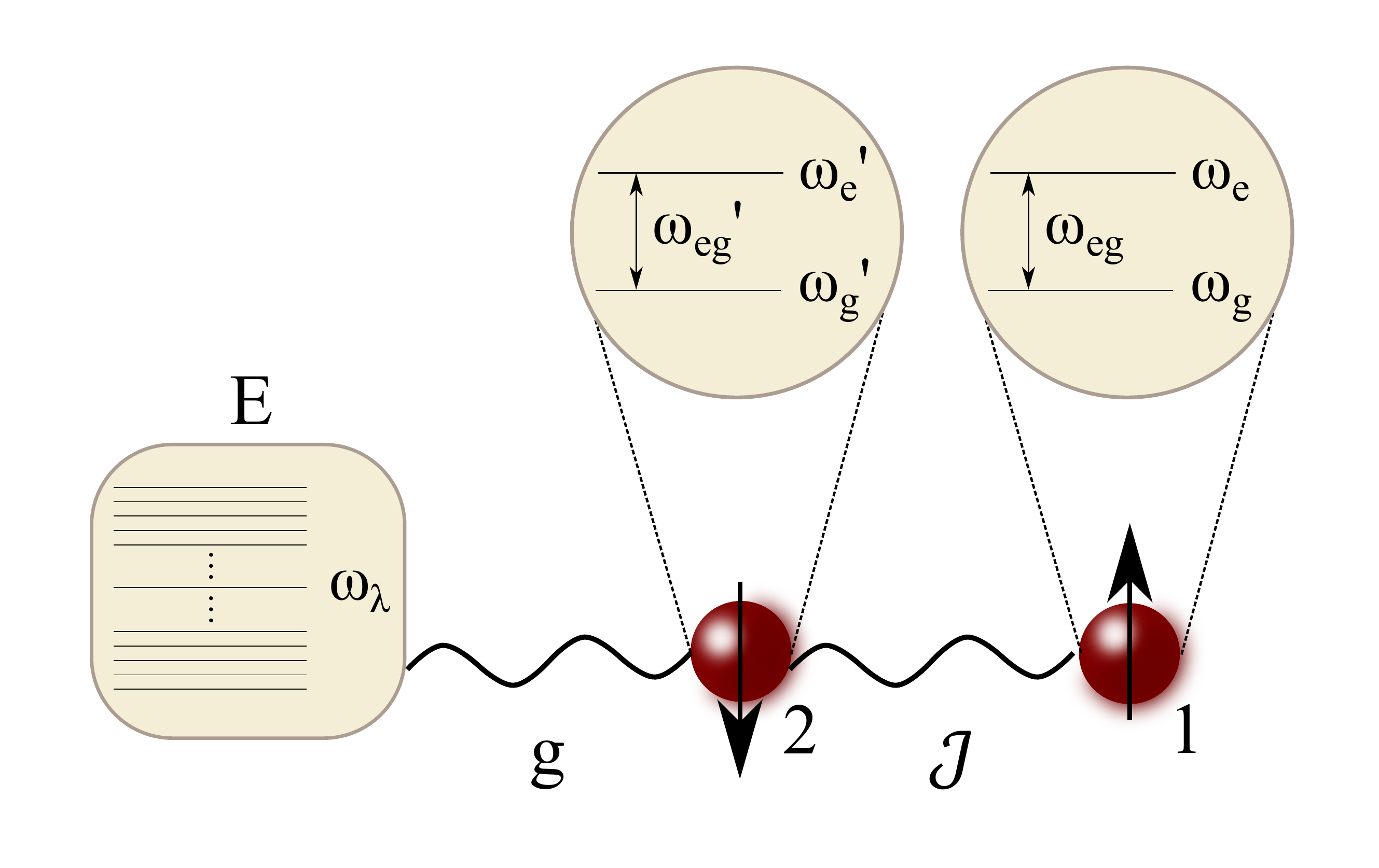}
		\caption[Fig1]{Schematic representation of the system at study. Two non-identical qubits are interacting with a coupling strength $\mathcal{J}$, while one of them is also coupled to an external environment $E$ via a coupling $g$.}
		\label{Fig1}
\end{figure}

By adopting the following transformations for the qubit and environment amplitudes; namely, $c_1 (t) = e^{-i \left( \omega_g' +  \omega_e  \right) t} \Tilde{c}_1 (t)$, $c_2 (t) = e^{-i \left( \omega_g +  \omega_e' \right)  t} \Tilde{c}_2 (t)$  and $ c_{\lambda}^{E}(t) = e^ {-i \left( \omega_g + \omega_g' + \omega_{\lambda} \right) t} \Tilde{c}_{\lambda}^{E}(t)$, it is easy to show that the time-dependent Schrodinger equation (TDSE) leads to the following equations of motion of the tilde amplitudes: 

\begin{subequations}
\beq
\frac{ d \Tilde{c}_1 (t)}{dt} =  - i \mathcal{J} \Tilde{c}_2 (t) e^{- i \varepsilon t},
\label{Amplitude_1}
\eeq

\beq
\frac{ d \Tilde{c}_2 (t)}{dt}  =  - i \mathcal{J} \Tilde{c}_1 (t) e^{+ i \varepsilon t} - i \sum_{\lambda} g\left( \omega_{\lambda} \right) e^{-i \Delta_{\lambda} t}  \Tilde{c}_{\lambda}^{E}(t), 
\label{Amplitude_2}
\eeq

\beq
\frac{ d \Tilde{c}_{\lambda}^{E}(t)}{dt} =  -i g\left( \omega_{\lambda} \right) e^{+i \Delta_{\lambda} t} \Tilde{c}_2 (t),
\label{Amplitude_E}
\eeq
\label{EqsofmotionIIA}
\end{subequations}
where $\varepsilon \equiv (\omega_e' - \omega_g') - (\omega_e - \omega_g) \equiv \omega_{eg}' - \omega_{eg}$ is the difference between the two qubit energies and $\Delta_{\lambda} \equiv \omega_{\lambda}- (\omega_e' - \omega_g') \equiv \omega_{\lambda} - \omega_{eg}'$ is the detuning between the energy of the ${\lambda}^{th}$ mode of the environment and the excitation energy of the second qubit.

Formal integration of Eqn. (\ref{Amplitude_E}) under the initial condition $\Tilde{c}_{\lambda}^{E}(0)=0$ (which is equivalent to $c_{\lambda}^{E}(0)=0$) and substitution back into Eqn. (\ref{Amplitude_2}), yields:

\beq
\begin{split}
\frac{ d \Tilde{c}_2 (t)}{dt}  = & - i \mathcal{J} \Tilde{c}_1 (t) e^{+ i \varepsilon t} \\
& - \int_0^t \sum_{\lambda} \left[ g \left( \omega_{\lambda} \right) \right]^2 e^{-i \Delta_{\lambda} \left( t - t' \right)}  \Tilde{c}_{2}(t') dt'.
\label{New_Amplitude_2}
\end{split}
\eeq
At this point we replace the sum over all the modes of the environment by a frequency integral, according to the relation $\sum_{\lambda} \left[  g (\omega_{\lambda}) \right]^2 \rightarrow \int d \omega J (\omega)$, where $J(\omega)$ is the spectral density of the environment. In view of this substitution, Eqn. (\ref{New_Amplitude_2}) becomes:

\beq
\frac{ d \Tilde{c}_2 (t)}{dt}  =  - i \mathcal{J} \Tilde{c}_1 (t) e^{+ i \varepsilon t} - \int_0^t R \left( t -t' \right)  \Tilde{c}_{2}(t') dt', 
\label{Eq_2}
\eeq
where $R(t)$ is defined via the following equation:

\beq
R (t) \equiv \int_0^{\infty} J (\omega) e^{- i \Delta   t } d\omega.
\label{R(t)}
\eeq
with $\Delta = \omega - \omega_{eg}'$. Eqns. (\ref{Amplitude_1}) and (\ref{Eq_2}) now form our set of differential equations we wish to solve for $\Tilde{c}_1 (t)$. Taking the Laplace transform of these Eqns. and using the Laplace transform properties of frequency shifting and convolution, we readily obtain:

\begin{subequations}
\beq
s F_1 (s) = c_1 (0) - i \mathcal{J} F_2 (s+i \varepsilon),
\label{Ltransform_1}
\eeq

\beq
s F_2 (s) = c_2 (0) - i \mathcal{J} F_1 (s-i \varepsilon) - B(s) F_2 (s),
\label{Ltransform_2}
\eeq
\end{subequations}
where $F_1 (s)$ and $F_2 (s)$ are the Laplace transforms of the the tilde amplitudes $\Tilde{c}_1 (t)$ and $\Tilde{c}_2 (t)$, respectively, while $B (s)$ is the Laplace transform of $R(t)$. Note that we also used the fact that the tilde amplitudes are equal to the amplitudes at $t=0$. Although the above set of equations can be solved for $F_1(s)$ and $F_2(s)$ for arbitrary initial conditions, for the purposes of our study we focus on the expression of $F_1(s)$ for initial excitation on the first qubit, i.e. $c_1(0)=1$ and $c_2(0)=0$. In that case, we can easily show that $F_1(s)$ is given by the following expression:
\beq
F_1(s) = \frac{1}{s+\frac{\mathcal{J}^2}{s + i \varepsilon + B \left( s + i \varepsilon \right)}}.
\label{F_1_final}
\eeq

In order to calculate the inversion integral to obtain the time dependence of $\Tilde{c}_1(t)$, we should first specify the spectral density function of the environment, so that we can derive $R(t)$ according to Eqn. (\ref{R(t)}) and hence the expression of its Laplace transform $B(s)$. Special cases of environments with Markovian, Lorentzian or Ohmic spectral densities are studied in great detail in the Results $\&$ Discussion section of our paper, revealing the regions of parameters that affect the onset of the Quantum Zeno regime.

It is important to note that our formalism can be used to explore much more complicated systems, involving arbitrary number of qubits and/or environments. A rather interesting result arises if we consider a system in which the qubit 1 of Fig. 1 does not interact directly with only one qubit (qubit 2) but with an arbitrary number of qubits $N$, each one of which is coupled to its own environment. Using our formulation we can show that, if all of the qubits are identical and the surrounding environments are characterized by the same spectral density, the Laplace transform of the tilde amplitude of the first qubit is given by:

\beq
F_1 (s)  = \frac{1}{s+   \frac{ N \mathcal{J}^2}{ s + B(s) }}.
\label{F_A_N_envs}
\eeq
This equation is essentially the same with Eqn. (\ref{F_1_final}) for $\varepsilon=0$ (identical qubits), with the exception of a factor of $N$ multiplying $\mathcal{J}^2$, where $N$ is the number of qubits interacting with qubit $1$. In other words, the system consisting of a qubit (qubit 1) interacting with $N$ qubits that communicate with $N$ respective environments with identical spectral densities can be effectively considered equivalent to a two-qubit + one environment system (Fig. 1) with a "collective coupling" $\sqrt{N} \mathcal{J}$ between the two qubits.

On the other hand, if all of the $N$ qubits that interact with qubit 1 are communicating with a common environment, it is straightforward to show that $F_1(s)$ acquires the form:
\beq
F_1 (s)  = \frac{1}{s+   \frac{ N \mathcal{J}^2}{ s + N B(s) }}.
\label{F_A_Common_env}
\eeq
where the factor of $N$ now multiplies both $\mathcal{J}^2$ and $B(s)$. 

\section{Results \& Discussion}

\subsection{Markovian environment}

By definition, a Markovian environment is an environment with a smooth, slowly varying spectral density as a function of the frequency. An attempt to find an expression of $R(t)$ in the extreme case of a totally flat spectral density $J( \omega ) = J = const.$ in Eqn. (\ref{R(t)}) leads to the problem of a divergent integral. One could therefore attempt to construct a spectral density that is not constant but changes very slowly as a function of $\omega$, such that the Markov approximation is valid. Such spectral density would enable the analytical derivation of $R(t)$ and hence its Laplace transform $B(s)$ which is necessary for the Laplace inversion of Eqn. (\ref{F_1_final}). Another solution would be to derive $R(t)$ and $B(s)$ for a Lorentzian spectral density and take the limit of the Lorentzian width becoming infinite, while keeping the ratio between the qubit-environment coupling strength and the width as constant, thus capturing the Markovian limit. However, without loss of rigor, the most practical way to take into account the effects of the coupling to the Markovian reservoir is to add a phenomenological decay rate and energy shift in the excited state energy of the second qubit, as has been well known theoretically for more than 50 years \cite{ref2}. For the sake of simplicity we neglect the energy shift, as it is of no essential relevance to our inquiry, and make the substitution $\omega_e' \rightarrow \omega_e' - i \Tilde{\gamma}/ 2$, which translates to $\varepsilon \rightarrow \varepsilon - i \Tilde{\gamma} / 2$, where $\Tilde{\gamma}$ is the decay rate of the second qubit excited energy, which also expresses the magnitude of the coupling $g$ between the second qubit and the Markovian reservoir. Substituting the expression $\varepsilon \rightarrow \varepsilon - i \Tilde{\gamma} / 2$ for $B(s)=0$ (the effects of the Markovian reservoir on the system are reflected only on changes of $\varepsilon$ that have been take into account phenomenologically) in Eqn. (\ref{F_1_final}) one finds that 

\beq
F_1(s) = \frac{s + ( \Tilde{\gamma}/ 2 + i \varepsilon) }{s^2 + ( \Tilde{\gamma}/ 2 + i \varepsilon) s + {\mathcal{J}}^2 }.
\label{F_1_Markovian}
\eeq

The Laplace inversion of Eqn. (\ref{F_1_Markovian}) provides us with the time evolution of the tilde amplitude of qubit 1. An insightful expression can be obtained in the case of identical qubits ($\varepsilon =0 $), where the Laplace inversion yields:
\beq
\begin{split}
\Tilde{c}_1(t)  & = e^{- \Tilde{\gamma} t /4} \Bigg[  \cos{ \left( \frac{t}{4} \sqrt{ {\left( 4 \mathcal{J} \right)}^2 - {\Tilde{\gamma}}^2} \right)} \\
& + \frac{\Tilde{\gamma}}{\sqrt{ {\left( 4 \mathcal{J} \right)}^2 - {\Tilde{\gamma}}^2} }  \sin{ \left( \frac{t}{4} \sqrt{ {\left( 4 \mathcal{J} \right)}^2 - {\Tilde{\gamma}}^2} \right)}  \Bigg] , \quad \Tilde{\gamma} \neq 4 \mathcal{J}
\end{split}
\label{c_1_Markovian_identical}
\eeq
and $\Tilde{c}_1(t) = e^{- \mathcal{J} t} \left( 1 + \mathcal{J}t \right)$, for $\Tilde{\gamma} = 4 \mathcal{J}$. It is interesting to observe that for $\Tilde{\gamma} \ll 4 \mathcal{J}$, the tilde amplitude of the first qubit follows an oscillatory behaviour with frequency equal to $\frac{\sqrt{ {\left( 4 \mathcal{J} \right)}^2 - {\Tilde{\gamma}}^2}}{4} $ along with an exponential decay. As $\Tilde{\gamma}$ approaches the value $4 \mathcal{J}$ the oscillations tend to disappear, and when $\Tilde{\gamma} = 4 \mathcal{J}$, the oscillatory part $\Bigg[  \cos{ \left( \frac{t}{4} \sqrt{ {\left( 4 \mathcal{J} \right)}^2 - {\Tilde{\gamma}}^2} \right)} + \frac{\Tilde{\gamma}}{\sqrt{ {\left( 4 \mathcal{J} \right)}^2 - {\Tilde{\gamma}}^2} }  \sin{ \left( \frac{t}{4} \sqrt{ {\left( 4 \mathcal{J} \right)}^2 - {\Tilde{\gamma}}^2} \right)}  \Bigg]$ reduces to a form which is linear on time, i.e. $(1 + \mathcal{J} t )$. In the $\Tilde{\gamma} \gg 4 \mathcal{J}$ limit it is straightforward to show that $\Tilde{c}_1(t) \rightarrow 1 $, using the identities $\cos{ \left( i x \right)} = \cosh{x}$ and $\sin{ \left( i x \right)} = i  \sinh{x}$. Although this result may seem counter-intuitive at first glance, it can be interpreted in terms of the QZE, i.e. a strong coupling between the second qubit and the Markovian environment causes the second qubit to freeze in its ground state, preventing qubit 1 from transferring population to qubit 2 and hence to the environment. Therefore the population of the first qubit becomes protected against dissipation. Note that for this scheme to work, it is crucial not to have an initially populated second qubit, because in that the case, the part of population of the second qubit would quickly dissipate due to the strong coupling between the latter and the environment.

Diagonalization of $\hat{\mathcal{H}}_S$ after the substitution $\omega_e' \rightarrow \omega_e' - i \Tilde{\gamma}/ 2$ leads to the following four eigenvalues: 

\begin{subequations}
\beq
\lambda_1 = \omega_g + \omega_g',
\eeq

\beq
\lambda_2 = \omega_e + \omega_e' - i \Tilde{\gamma}/2, 
\eeq

\beq
\begin{split}
\lambda_3 = & \frac{1}{2} \left( \omega_g + \omega_g' + \omega_e + \omega_e' - i \Tilde{\gamma}/2 \right) \\
& - \frac{1}{4} \sqrt{ {\left( 4 \mathcal{J} \right)}^2 - \left({\Tilde{\gamma}} + 2i \varepsilon \right)^2},
\end{split}
\eeq

\beq
\begin{split}
\lambda_4 = & \frac{1}{2} \left( \omega_g + \omega_g' + \omega_e + \omega_e' - i \Tilde{\gamma}/2 \right)  \\
& + \frac{1}{4} \sqrt{ {\left( 4 \mathcal{J} \right)}^2 - \left({\Tilde{\gamma}} + 2i \varepsilon \right)^2}.
\end{split}
\eeq

\end{subequations}

\begin{figure}[t] 
	\centering
	\includegraphics[width=8.65cm]{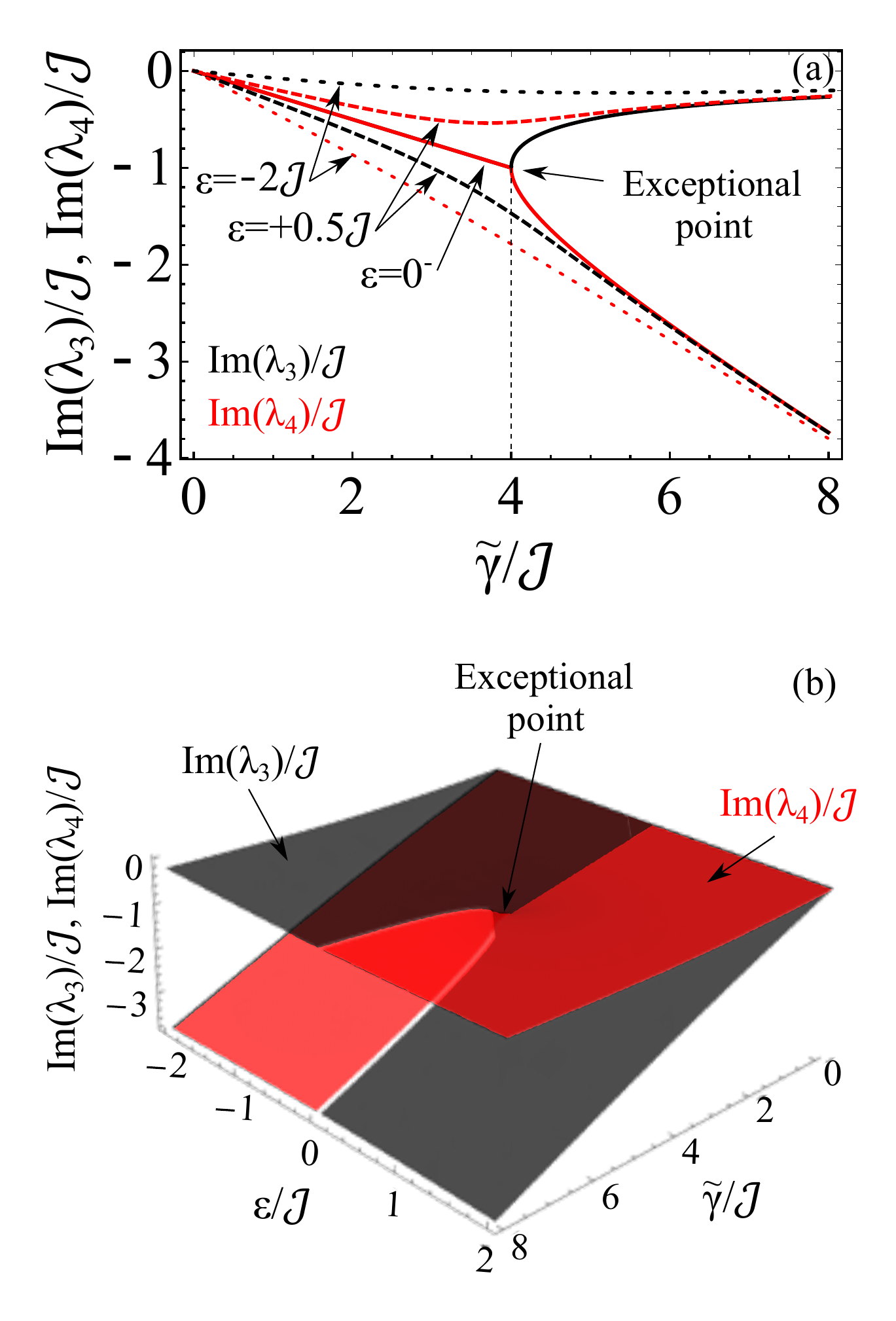}
		\caption[Fig2]{(a) Imaginary parts of the eigenvalues $\lambda_3$ (black) and $\lambda_4$ (red) as a function of $\Tilde{\gamma}$ for various values of the energy difference between the two qubits $\varepsilon$. Solid line: $\varepsilon=0$, dashed line: $\varepsilon=-2 \mathcal{J}$, dotted line: $\varepsilon= 0.5 \mathcal{J}$. The vertical dashed line indicates the position of the exceptional point for $\varepsilon=0^{-}$, i.e. $\Tilde{\gamma}= 4 \mathcal{J}$. (b) Imaginary parts of the eigenvalues $\lambda_3$ (black) and $\lambda_4$ (red) in the $\varepsilon$ - $\gamma$ plane.}
		\label{Fig2}
\end{figure}

We should note that since we focus on the single-excitation subspace we only need to consider the eigenvalues $\lambda_3$ and $\lambda_4$. Equivalently, if we diagonalized $\hat{\mathcal{H}}_S$ in the single-excitation subspace, which would be essentially a $2 \times 2$ matrix, its eigenvalues are $\lambda_3$ and $\lambda_4$.

In Fig. 2(a) we plot the imaginary part of the eigenvalues $\lambda_3$ and $\lambda_4$ as a function of $\Tilde{\gamma}$ for various values of the energy difference between the two qubits $\varepsilon$. As shown, in the case of identical qubits ($\varepsilon=0$), the imaginary parts of the two eigenvalues coalesce up to $\Tilde{\gamma}=4 \mathcal{J}$ and they split for $\Tilde{\gamma}>4 \mathcal{J}$. The point $\Tilde{\gamma}=4 \mathcal{J}$ is called exceptional point \cite{ref39,ref40} and marks the boundary between the unbroken and the broken $\mathcal{P} \mathcal{T}$ symmetry of the Hamiltonian. At the exceptional point, both the real and the imaginary parts of $\lambda_3$ and $\lambda_4$ coalesce, while the corresponding eigenvectors $\ket{\phi_3}$ and $\ket{\phi_4}$ given by the expressions below, become parallel.

\begin{figure*}[t] 
	\centering
	\includegraphics[width=17.5cm]{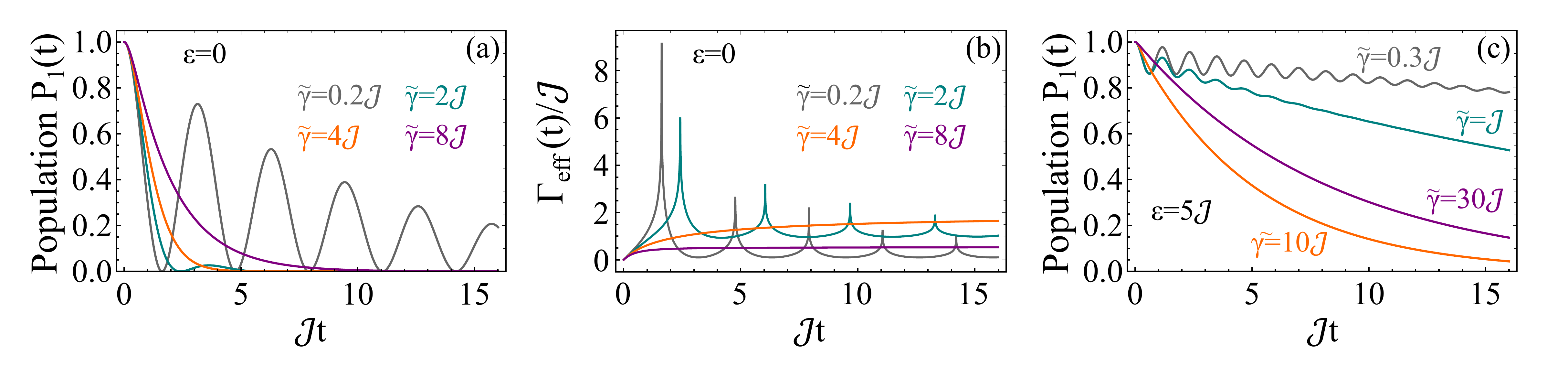}
		\caption[Fig3]{(a) Population of qubit 1 in the configuration of Fig. 1, as a function of the time for various values of $\Tilde{\gamma}$ and $\varepsilon=0$. Gray line: $\Tilde{\gamma}=0.2 \mathcal{J}$, teal line: $\Tilde{\gamma}=2 \mathcal{J}$, orange line: $\Tilde{\gamma}=4 \mathcal{J}$ and purple line: $\Tilde{\gamma}=8 \mathcal{J}$. (b) Time dynamics of the effective decay $\Gamma_{\text{eff}}$ using the same parameters with those of panel (a). (c) Population of qubit 1, as a function of the time for various values of $\Tilde{\gamma}$ and $\varepsilon=5\mathcal{J}$. Gray line: $\Tilde{\gamma}=0.3 \mathcal{J}$, teal line: $\Tilde{\gamma}= \mathcal{J}$, orange line: $\Tilde{\gamma}=10 \mathcal{J}$ and purple line: $\Tilde{\gamma}=30 \mathcal{J}$.}
		\label{Fig3}
\end{figure*}

\begin{subequations}
\beq
\begin{split}
\ket{\phi_3} =  \ket{g}_1 \ket{e}_2 + &  \frac{1}{2 \mathcal{J}}  \Bigg[  - \left( \varepsilon - i \Tilde{\gamma}/ 2 \right) \\
& - \frac{1}{2} \sqrt{ {\left( 4 \mathcal{J} \right)}^2 - \left({\Tilde{\gamma}} + 2i \varepsilon \right)^2} \Bigg] \ket{e}_1 \ket{g}_2,
\end{split}
\eeq

\beq
\begin{split}
\ket{\phi_4} =  \ket{g}_1 \ket{e}_2  + &  \frac{1}{2 \mathcal{J}}  \Bigg[  -  \left( \varepsilon - i \Tilde{\gamma}/ 2 \right) \\
& + \frac{1}{2} \sqrt{ {\left( 4 \mathcal{J} \right)}^2 - \left({\Tilde{\gamma}} + 2i \varepsilon \right)^2} \Bigg] \ket{e}_1 \ket{g}_2.
\end{split}
\eeq
\end{subequations}

For $\varepsilon \neq 0$, the imaginary parts of $\lambda_3$ and $\lambda_4$ are different for any value of $\Tilde{\gamma} \neq 0$, therefore no exceptional point exists in that case. It is interesting to note that there is an abrupt interchange between the values of $\Im(\lambda_3)$ and $\Im(\lambda_4)$ as $\varepsilon$ approaches zero. This behaviour can be explained by visualizing these quantities in the $\varepsilon$ - $\gamma$ plane as in Fig. 2(b). This abrupt change is based on the complex eigenvalue topology of the involved intersecting Riemann sheets across the $\varepsilon=0$ surface. The particular topology also suggests that the directionality of the motion in the $\varepsilon$ - $\gamma$ space, when encircling the exceptional point with starting points on different Riemann sheets, plays an important role on the final outcome.

In Fig. 3(a) we show the time population dynamics of the first qubit of the configuration depicted in Fig. 1. The population dynamics are studied for various values of $\Tilde{\gamma}$ and $\varepsilon=0$ (identical qubits). For small values of $\Tilde{\gamma}$ such that $\Tilde{\gamma} \ll 4\mathcal{J}$ (gray line), the population exhibits damped oscillations indicative of the transfer of the excitation to the second qubit along with the environmental dissipation. As $\Tilde{\gamma}$ is increased the oscillations are becoming increasingly damped and less frequent (teal line), while at the critical coupling $\Tilde{\gamma} = 4\mathcal{J}$ where the exceptional point lies, the oscillations disappear and the population dynamics are given by the expression $P_1 (t) = e^{- 2 \mathcal{J} t} \left( 1 + \mathcal{J} t \right)^2$ (orange line). For increasing values of $\Tilde{\gamma}$ the population retains its non-oscillatory behaviour and becomes increasingly protected against dissipation through the QZE (purple line). In the limit $\Tilde{\gamma} \gg 4\mathcal{J}$, as also our analytical study suggests, the population of the first qubit remains essentially frozen in its initial value. Strictly speaking complete freezing occurs in the limit of infinite $\tilde{\gamma}$, which is of only mathematical interest, as it is the range of finite but large values, in the sense of the above inequality, that are of realistic relevance.

The same conclusions can be deduced by studying the effective decay rate of the probability $P_1(t)$, an important and widely used tool in the context of QZE in open quantum systems, defined as:

\beq
\Gamma_{\text{eff}} (t) \equiv - \frac{1}{t} \ln [P_1 (t)].
\eeq

In Fig. 3(b) we show the time dynamics of the effective decay rate $\Gamma_{\text{eff}} (t)$ using parameters identical to those of Fig. 3(a). As becomes evident, the effective decay rate for $\Tilde{\gamma} < 4\mathcal{J}$ exhibits peaks associated with the population oscillations between the two qubits, whose frequency depends upon $\Tilde{\gamma}$. The time between subsequent peaks increases, as $\Tilde{\gamma}$ is increased, while each subsequent peak is less pronounced compared to the preceding one. For $\Tilde{\gamma} \geq 4\mathcal{J}$, the effective decay rate does not exhibit any peak, and its value tends to decrease as $\Tilde{\gamma}$ is increased.

\begin{figure}[t] 
	\centering
	\includegraphics[width=7cm]{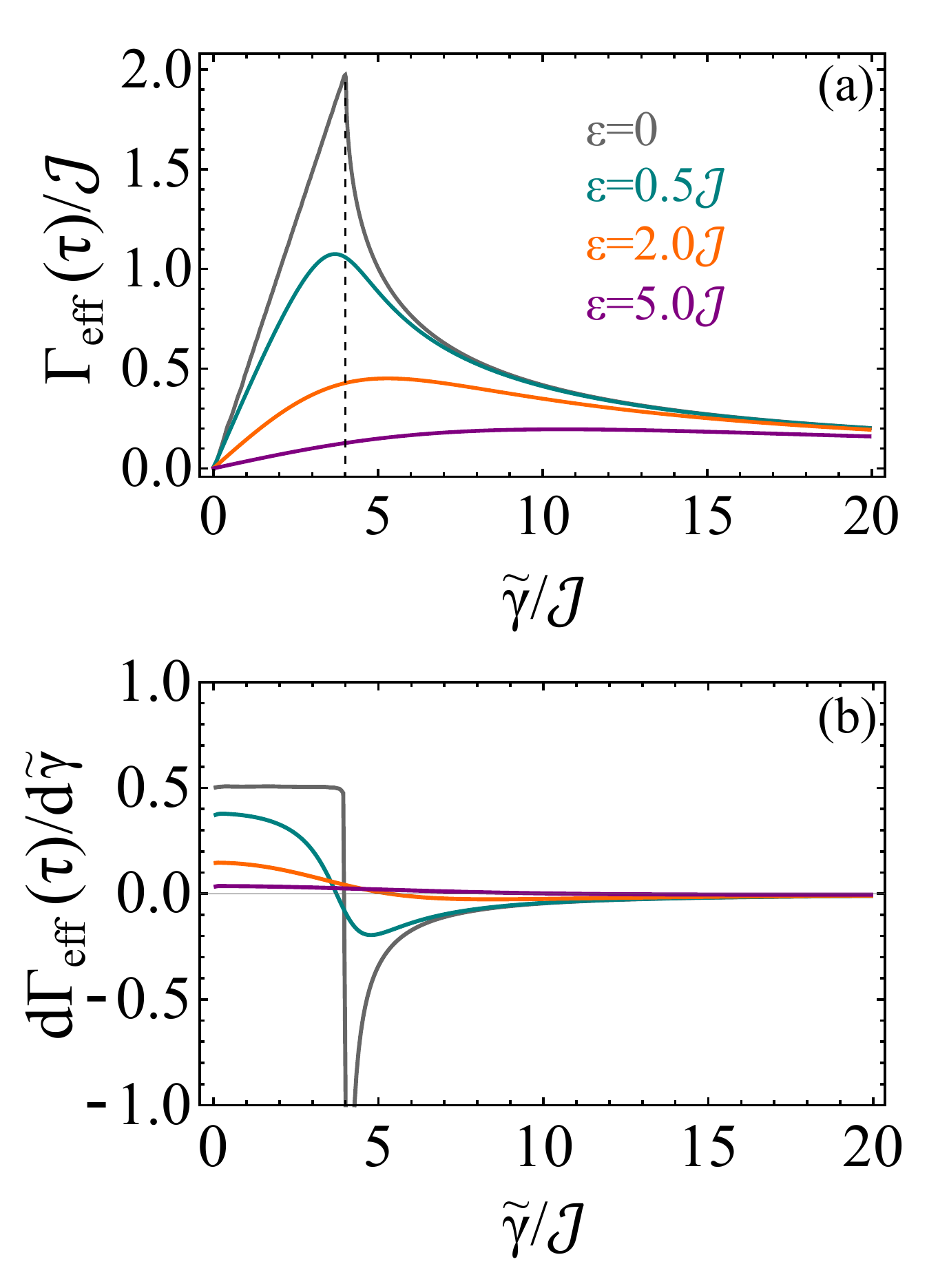}
		\caption[Fig4]{(a) Effective decay rate for times $\tau$ much larger than any other timescale of the system, as function of $\Tilde{\gamma}$, for various qubit energy differences $\varepsilon$. The vertical dashed line at $\Tilde{\gamma}=4 \mathcal{J}$ indicates the position of the peak for $\varepsilon=0$. Gray line: $\varepsilon=0$, teal line: $\varepsilon=0.5 \mathcal{J}$, orange line: $\varepsilon=2.0 \mathcal{J}$ and purple line: $\varepsilon= 5.0 \mathcal{J}$. (b) Derivative of the effective decay rate for times $\tau$ much larger than any other timescale of the system, with respect to $\Tilde{\gamma}$, as a function of $\Tilde{\gamma}$. The parameters are the same with those of panel (a).}
		\label{Fig4}
\end{figure}

\begin{figure*}[t] 
	\centering
	\includegraphics[width=17.8cm]{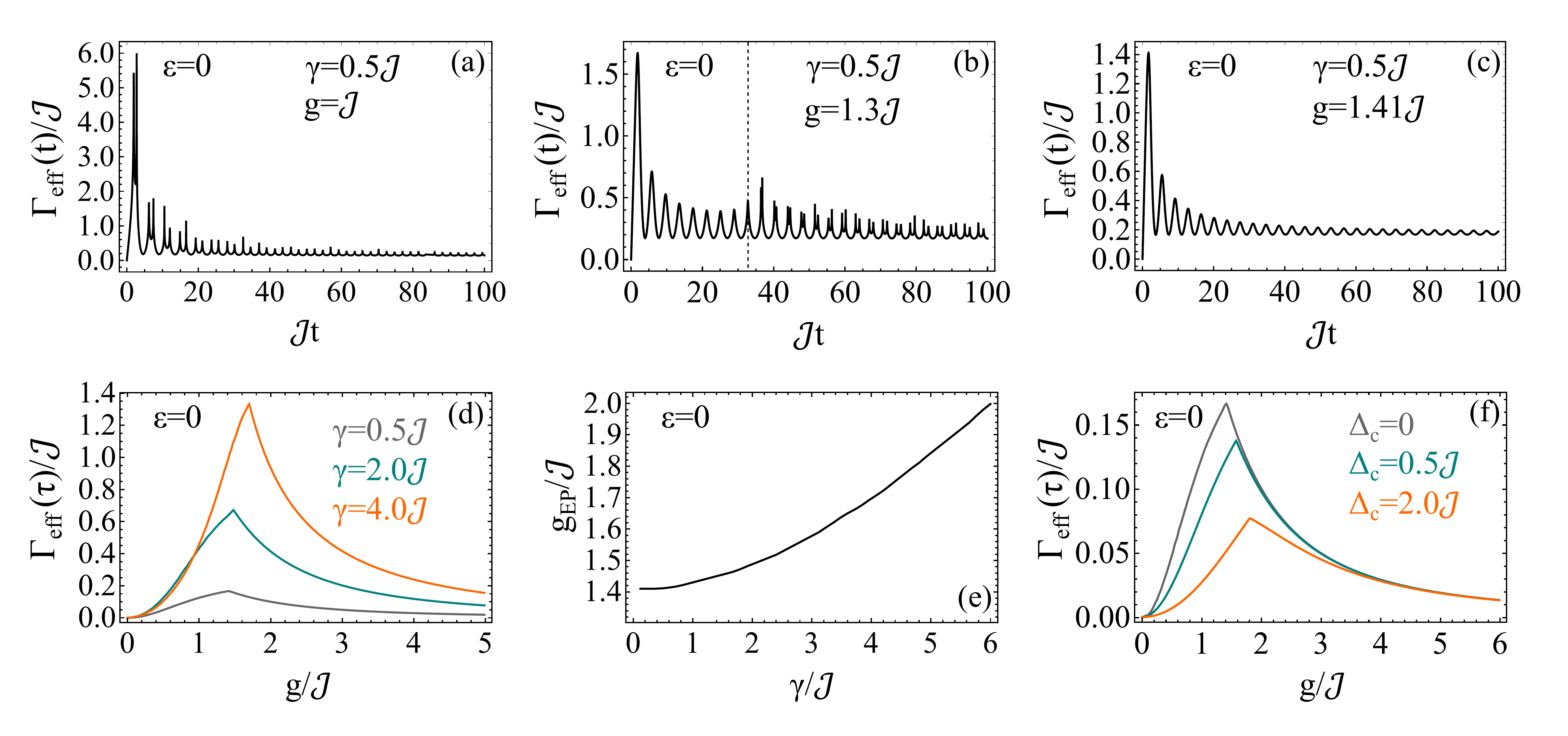}
		\caption[Fig5]{(a) Time dynamics of the effective decay rate for $\varepsilon=0$ (identical qubits) and a Lorentzian boundary environment with parameters $\gamma=0.5 \mathcal{J}$, $g= \mathcal{J}$ and $\Delta_c=0$. (b) Same as panel (a) but for $g= 1.3\mathcal{J}$. The vertical dashed line indicates the time up to which the decay rate oscillations are smooth. (c) Same as panel (a) but for $g= 1.41\mathcal{J}$. (d) Effective decay rate for times $\tau$ much larger than any other timescale of the system, as function of $g$, for various Lorentzian widths $\gamma$, $\varepsilon=0$ and $\Delta_c=0$. (e) Location of the exceptional point $g_{\text{EP}}$ as a function of $\gamma$ for $\varepsilon=0$ and $\Delta_c=0$. (f) Effective decay rate for times $\tau$ much larger than any other timescale of the system, as function of $g$, for various values of the detuning $\Delta_c$ between the Lorentzian peak and the qubit frequency of the second qubit. The parameters used are: $\epsilon=0$ and $\gamma = 0.5\mathcal{J}$. Gray line: $\Delta_c=0$, teal line: $\Delta_c=0.5 \mathcal{J}$ and orange line: $\Delta_c=2.0 \mathcal{J}$.}
		\label{Fig5}
\end{figure*}

However, the situation is quite  different if the two qubits  non-identical, as is the case of Fig. 3(c) where $\varepsilon= 5 \mathcal{J}$. As also Fig. 3(c) suggests, we do not expect any qualitative transition on the system's response as $\Tilde{\gamma}$ approaches and exceeds the value $\Tilde{\gamma}=4 \mathcal{J}$. In this case, what determines the dissipation behaviour is the ratio between $\varepsilon$ and $\Tilde{\gamma}$. In other words, as $\Tilde{\gamma}$ is increased the dissipation of $P_1(t)$ is also increased.  But if $\Tilde{\gamma}$ becomes sufficiently larger than $\varepsilon$, the picture changes with the population of the first qubit becoming increasingly robust against dissipation. Again, even for $\varepsilon \neq 0$, in the limit where $\Tilde{\gamma}$ is much bigger than $\mathcal{J}$ and $\varepsilon$, the QZE freezes the dynamics of the second qubit, inducing thus a hindering of the decay of the first qubit population.

As also evident from Fig. 3(b), in the long-time limit, the effective decay rate $\Gamma_{\text{eff}} (t)$ tends to stabilize to a finite non-zero value. The results of Fig. 3(c) become much clearer if we plot $\Gamma_{\text{eff}} (\tau)$ as a function of $\Tilde{\gamma}$, where $\tau$ is defined to be a time much larger than any other timescale of the system. This quantity informs us about the onset of the QZE since it indicates the coupling $\Tilde{\gamma}$ where the decay becomes maximum and decreases thereafter. As seen in Fig. 4(a), for $\varepsilon=0$ and increasing $\Tilde{\gamma}$, the decay is also increased until the position $\Tilde{\gamma}=4\mathcal{J}$ where it becomes maximum. As $\Tilde{\gamma}$ crosses the value $4\mathcal{J}$ there is an abrupt change in the behaviour of the decay as also suggested by Fig. 4(b) where we show the derivative of the effective decay rate with respect to $\Tilde{\gamma}$. In view of the analysis of Fig. 2 the sharp peak at $\Tilde{\gamma}=4\mathcal{J}$ is therefore indicating the position of the exceptional point.

On the other hand, for finite $\varepsilon$, i.e. non-identical qubits, although the effective decay rate calculated at large times also exhibits a maximum, the curve around the maximum is smooth. In this case, according to Fig. 2, we should not expect any exceptional points at any coupling strength $\Tilde{\gamma}$. For increasing $\varepsilon$, the decay rate as function of $\Tilde{\gamma}$ exhibits an increased width which indicates that it takes a larger coupling window to cross the maximum and move from regions of increasing dissipation to the Quantum Zeno regime, in compatibility with Fig. 3(c). 

Our results suggest that the onset of the QZE is not -necessarily- associated with a presence of an exceptional point but with a peaked structure of the effective decay rate as a function of the coupling strength between the second qubit and the environment. The presence of an exceptional point on the other hand always indicates an abrupt phase transition from the anti-Zeno to the Zeno regime and it is associated with a sharp peak of the effective decay rate as a function of $\Tilde{\gamma}$. The link between these sharp peaks and the presence of EPs has also been pointed out in a recent paper by P. Kumar et al. \cite{ref34}. The sharpness of the peak can be easily identified trough discontinuities of the first derivative of the effective decay rate with respect to  $\Tilde{\gamma}$ as in Fig. 4(b). Therefore, if one can measure the quantity $\Gamma_{\text{eff}} (\tau)$ as a function of $\Tilde{\gamma}$ for $\tau$ much larger than any other timescale of the system, he/she could argue about the presence of EPs by just studying its peak structure. As will become evident from what follows, this method appears to be very useful in cases of systems where explicit expressions of effective Hamiltonians do not exist and therefore no diagonalization is possible.

\subsection{Lorentzian environment}

In order to find the time dependence of the tilde amplitude $\Tilde{c}_1(t)$ for a Lorentzian boundary environment, one needs to calculate the function $R(t)$ via Eq. (\ref{R(t)}) for a Lorentzian spectral density $J \left( \omega \right)$ and find its Laplace transform $B(s)$, necessary for the inversion of the Laplace transform $F_1(s)$. In a recent paper \cite{ref1}, using similar notation, we calculated $B(s)$ for a Lorentzian spectral density distribution with positive peak frequency and negligible extension to negative frequencies and showed that

\beq
B (s) = \frac{g^2}{s+ \frac{\gamma}{2} + i \Delta_c},
\label{B(s)_Lorentzian}
\eeq
where, $g$ is the coupling strength constant between the second qubit and the environment, $\gamma$ is the width of the distribution and $\Delta_c \equiv \omega_c - \omega_{eg}'$ is the detuning between the peak frequency $\omega_c$ of the distribution and the qubit frequency $\omega_{eg}'$ of the second qubit. Substitution of Eq. (\ref{B(s)_Lorentzian}) back to Eq. (\ref{F_1_final}) leads to an expression involving a third order polynomial with respect to $s$ in the denominator. Although the Laplace inversion can be carried out analytically, the resulting expression of $\Tilde{c}_1(t)$ is too lengthy to be insightful.

In contrast to the previous case, where the environment was characterized by a  Markovian spectral density, now it is not possible to develop an effective Hamiltonian characterizing the open quantum system by eliminating the reservoir degrees of freedom. This inability is associated with the non-Markovian character of the Lorentzian spectral density, which enables information exchange between the system and the environment within finite times. Therefore an attempt to find the eigenergies of the open system as a probe of its exceptional points seems ineffectual. Based, however, on the results of the previous subsection, deduced from the study of the effective decay rate maxima, in comparison to what we know from the spectrum of the non-Hermitian Hamiltonian and its exceptional points, we can track the EPs of the system damped by a Lorentzian reservoir.

In Fig. 5(a) we plot the effective decay rate of the first qubit as a function of the time for $\varepsilon=0$, $\gamma=0.5\mathcal{J}$, $\Delta_c=0$ and $g=\mathcal{J}$. Based on the form of the peaked structure of $\Gamma_{\text{eff}} (\tau)$ as a function of $g$ for $\tau$ much larger than any other timescale of the system, we expect an exceptional point at $g=1.41 \mathcal{J}$ (see Fig. 5(d), gray line). For $g$ smaller than the position of the exceptional point, which we will denote by $g_{\text{EP}}$ hereafter, the effective decay rate as a function of the time exhibits sharp peaks indicative of the transfer of populations between the two qubits. Note that, contrary to the Markovian environment case, part of the excitation can now be transferred from the open system to the environment and vice versa, within finite times. As $g$ is increased towards the value $g_{\text{EP}}$, the sharp peaks begin to be gradually substituted by smooth oscillations (Fig. 5(b)) up to the point where $g=g_{\text{EP}}$ at which the $\Gamma_{\text{eff}} (t)$ dynamics exhibit only smooth oscillations, as in Fig. 5(c). Note that for $g \geq g_{\text{EP}}$, the system lies in the region where the QZE starts to inhibit the evolution of the second qubit. As a result, as $g$ increases, the population of the second qubit becomes increasingly negligible and the smooth oscillations of the effective decay rate of the first qubit reflect oscillations directly between the latter and the environment.

\begin{figure*}[t] 
	\centering
	\includegraphics[width=17.8cm]{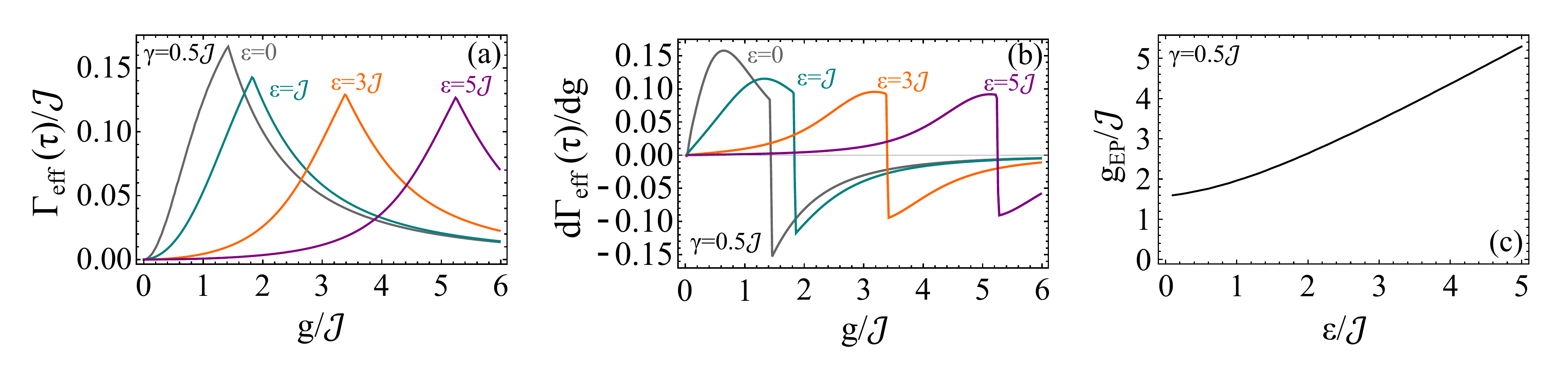}
		\caption[Fig6]{(a) Effective decay rate for times $\tau$ much larger than any other timescale of the system, as function of $g$, for various energy differences $\varepsilon$ between the two qubits, $\gamma = 0.5 \mathcal{J}$ and $\Delta_c=0$ (Lorentzian boundary reservoir). Gray line: $\gamma = 0$, teal line: $\gamma= \mathcal{J}$, orange line: $\gamma = 3 \mathcal{J}$ and purple line: $\gamma = 5 \mathcal{J}$. (b) Derivative of the effective decay rate for times $\tau$ much larger than any other timescale of the system, with respect to g, as a function of g, for various $\varepsilon$. The values of the parameters are chosen the same as with panel (a). (c) Location of the exceptional point as a function of $\varepsilon$ for $\gamma = 0.5 \mathcal{J}$ and $\Delta_c=0$.}
		\label{Fig6}
\end{figure*}

In Fig. 5(d) we show how the effective decay rate at long times $\tau$ behaves as a function of $g$ for various Lorentzian widths and $\varepsilon=0$. As expected on physical grounds, the decay rate is overall increased as $\gamma$ increases. At the same time the sharp peak of the curve which indicates the position of the exceptional point, moves towards larger $g$. The dependence between the position of the exceptional point $g_{\text{EP}}$ and the Lorentzian width $\gamma$, is depicted in Fig. 5(e). At the same time the maximum of the curve is also affected by the value of detuning between the Lorentzian peak and the qubit frequency of the second qubit (Fig. 5(f)). In view of the above results, we can confidently argue that the positions of the exceptional points in the case of a Lorentzian reservoir show great sensitivity to the values of the Lorentzian parameters $\gamma$ and $\Delta_c$.

For the results of Fig. 5 we have assumed that $\varepsilon=0$, i.e. the two identical qubits. In Fig. 6 we examine the effects of a non-zero energy difference between the two qubits on the QZE onset for Lorentzian reservoirs. In Fig. 6(a) we plot the effective decay rate at long times $\tau$ as a function of the qubit-environment coupling strength $g$ for various values of $\varepsilon$. As $\varepsilon$ is increased, the position of the maximum of the curve tends towards larger coupling strengths, as was the case for a Markovian-damped open system (see Fig. 4). There are however two striking differences. First, in the Markovian case the value of $\varepsilon$ affected significantly the width of the effective decay curve whereas for Lorentzian reservoirs, the increase of $\varepsilon$ results roughly to a displacement of the curve towards larger $g$. Second and most important, in the Markovian case, for any value of $\varepsilon$, the effective decay rate exhibited a smooth maximum, except the case $\varepsilon=0$ where the peak was sharp (see derivative in Fig. 4(b)), and was marked by the presence of an exceptional point. On the other hand, for a Lorentzian boundary reservoir, the peak of the effective decay rate is sharp for any value of $\varepsilon$. This result can be verified upon inspection of the discontinuities of the effective decay rate derivative with respect to $g$ as a function of $g$ for various values of $\varepsilon$ (Fig. 6(b)). Therefore, the maxima points of the effective decay rate mark the existence of exceptional points for any value of $\varepsilon$ in the case of a Lorentzian boundary reservoir. The dependence of the positions of such points as a function of $\varepsilon$ is depicted in Fig. 6(c).

\subsection{Ohmic environment}
In this subsection, we examine the case of a boundary environment  characterized by an Ohmic spectral density \cite{ref47,ref48}. In that case $J(\omega)$ is given by:
\beq
J(\omega) = \mathcal{N} g^2 \omega_c \left( \frac{\omega}{\omega_c} \right)^{\mathcal{S}} \exp \left(- \frac{\omega}{\omega_c} \right),
\label{OhmicSD}
\eeq
where $\omega_c$ is the so-called Ohmic cut-off frequency and $\mathcal{S}$  the Ohmic parameter, characterizing whether the spectrum of the reservoir is sub-Ohmic ($\mathcal{S} < 1$), Ohmic ($\mathcal{S}$ = 1) or super-Ohmic ($\mathcal{S} > 1$). $\mathcal{N}$ is a normalization constant given by the relation $\mathcal{N}= \frac{1}{ \left( {\omega_c} \right)^2 \Gamma \left( 1 + \mathcal{S} \right)}$, where $\Gamma(z)$ is the gamma function. 

\begin{figure*}[t] 
	\centering
	\includegraphics[width=18.2cm]{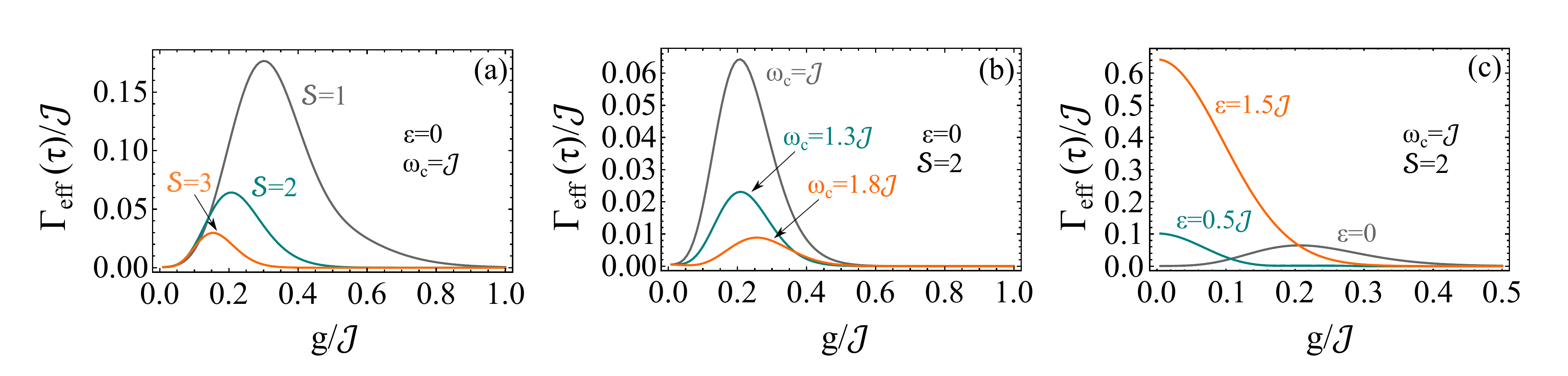}
		\caption[Fig7]{(a) Effective decay rate for times $\tau$ much larger than any other timescale of the system, as function of $g$, for various Ohmic parameters $\mathcal{S}$, $\omega_c=\mathcal{J}$, $\varepsilon=0$ and $\omega_{eg}=6 \mathcal{J}$. Gray line: $\mathcal{S}=1$, teal line: $\mathcal{S}=2$, orange line: $\mathcal{S}=3$. (b) Effective decay rate for times $\tau$ much larger than other timescale of the system, as function of $g$, for various cut-off frequencies $\omega_c$, $\mathcal{S}=2$, $\varepsilon=0$ and $\omega_{eg}=6 \mathcal{J}$. Gray line: $\omega_c=\mathcal{J}$, teal line: $\omega_c=1.3\mathcal{J}$, orange line: $\omega_c=1.8\mathcal{J}$. (c) Effective decay rate for times $\tau$ much larger than other timescale of the system, as function of $g$, for various energy differences $\varepsilon$ between the two qubits, $\mathcal{S}=2$, $\omega_c=\mathcal{J}$ and $\omega_{eg}=6 \mathcal{J}$. Gray line: $\varepsilon=0$, teal line: $\varepsilon=0.5\mathcal{J}$, orange line: $\varepsilon=1.5\mathcal{J}$.  }
		\label{Fig7}
\end{figure*}

The corresponding function $B(s)$ which is the Laplace transform of the function $R(t)$ given by Eq. (\ref{R(t)}), has been calculated in a previous work for Ohmic spectral densities \cite{ref1} and was found to be given by:
\beq
B(s) = - g^2 \frac{i^{1- \mathcal{S}}}{\omega_c}   e^{-i K(s)}  \left[ K(s) \right]^{\mathcal{S}} \Gamma \left(- \mathcal{S}, - i K(s) \right), 
\label{B_Ohmic}
\eeq
where $K(s) \equiv \left( s-i \omega_{eg} \right) / \omega_c$ and $\Gamma(a,z)$ is the incomplete gamma function. Substitution of Eq. (\ref{B_Ohmic}) back to Eq. (\ref{F_1_final}) leads to an expression of $F_1(s)$ whose inverse Laplace transform can be calculated numerically to yield the time dependence of $\Tilde{c}_1(t)$.

In Fig. 7 we plot the effective decay rate of the first qubit at long times $\tau$, as a function of the qubit-environment coupling strength $g$, for various combinations of the remaining parameters. In Fig. 7(a) we examine the effects of varying the Ohmic parameter $\mathcal{S}$ on the behaviour of the effective decay rate profile in the case of identical qubits ($\varepsilon=0)$.  The effective decay rate is now found to exhibit a peak for any value of $\mathcal{S}$.  But it is not sharp, i.e. the first derivative of the effective decay rate with respect to $g$, as a function of $g$, does not exhibit a discontinuity at the position of the peak. Although this suggests that the QZE occurs for any value of $\mathcal{S}$, it is not accompanied by the presence of an EP. The onset of the QZE (position of the maximum) decreases as the Ohmic parameter is increased. At the same time, the overall decay rate decreases as $\mathcal{S}$ is increased, which can be interpreted in terms of the form of the Ohmic spectral density distribution as a function of $\mathcal{S}$. In particular, for fixed $\omega_c$ and increasing $\mathcal{S}$, the distribution tends to flatten, causing more dominant modes of the distribution to be off-resonance from the qubit frequency, thus damping the system less efficiently. The difference between the effects of the Ohmic and Lorentzian distributions can be attributed to the fact that the flattening of the distribution and the position of its peak is controlled by different parameters in the two cases. For the Lorentzian, they are $\gamma$ and $\Delta_c$ (for fixed $\omega_{eg}$), respectively, whereas for an Ohmic distribution, both of them depend on the Ohmic parameters $\mathcal{S}$ and $\omega_c$. Note that the Ohmic distribution exhibits a peak at the frequency $\mathcal{S} \omega_c$.

In Fig. 7(b) we have kept the Ohmic parameter fixed to the value $\mathcal{S}=2$ and examined the behaviour of the effective decay rate, as a function of the cut-off frequency of the distribution. The results indicate that the onset of the QZE occurs for larger qubit-environment couplings $g$, as the cut-off frequency is increased. Again, the effective decay rate shows no evidence for the presence of EPs, for any combination of the Ohmic parameters. The values of the decay rate decrease, as $\omega_c$ is increased, owing to the flattening of the distribution for fixed $\mathcal{S}$ and increasing $\omega_c$, along the lines of interpretation in the previous paragraph.

Finally, in Fig. 7(c) we examine the behaviour of the effective decay rate for various values of the energy difference between the two qubits. Interestingly, contrary to the case of identical qubits, when $\varepsilon \neq 0$, the effective decay rate is maximum at $g=0$ and decreases as $g$ is increased. This result indicates that the system lies in the QZE regime for any value of $g$.

\section{Concluding Remarks}

We have investigated a method of tracking EPs in a non-Markovian open quantum system, for which a closed form expression of the effective Hamiltonian describing the open system may not exist. In that case, the EPs of the system cannot be found by following the usual procedure of Hamiltonian diagonalization, as would have been the case for a quantum system damped by one or more Markovian reservoirs. Although our theory in this paper deals with the simple case of two non-identical qubits, one of which interacts with a reservoir of arbitrary spectral density, our method is readily generalizable to any number of qubits and reservoirs.

The method is based upon studying the behaviour of the effective decay rate of the first qubit as a function of the coupling between the environment and the second qubit. We first studied the case of Markovian damping where the system is diagonalizable and we compared the effective decay rate analysis with the analysis in terms of the eigenvalues/eigenergies of the open system. The results indicated that although a peak of the effective decay rate term is always associated with the onset of the QZE, if the peak is sharp (i.e. if the first derivative of the effective decay rate with respect to $g$, as a function of $g$ exhibits a discontinuity), the system has an EP at the position of the peak. 

We have further examined the cases of boundary environment characterized by  Lorentzian, as well as Ohmic spectral densities. For Lorentzians, we have shown that the system will always have a single EP, for any combination of the parameters of the spectral density, i.e. its width and the detuning of its peak from the qubit frequency. The position of the EP ($g_{EP}$) has been found to shift towards higher values as $\gamma$ increases for fixed $\Delta_c$ or as $\Delta_c$ increases for fixed $\gamma$. Interestingly, in contrast to the case of Markovian damping where the EP exists only for identical qubits ($\varepsilon=0$), for a Lorentzian reservoir, an EP is always present irrespective of the value of $\varepsilon$. On the other hand, for reservoirs with an Ohmic spectral density, our results indicate that, although the system has a critical coupling marking the onset of the QZE (peak of the effective decay rate curve), this onset is not accompanied by the presence of an EP (i.e. the peak is not sharp), for any combination of the Ohmic parameters. The position of this onset was found to move towards smaller values of $g$, as the Ohmic parameter $\mathcal{S}$ is increased or as the Ohmic cut-off frequency $\omega_c$ is decreased. For $\varepsilon \neq 0$ the effective decay rate is maximum at $g=0$, decreasing  monotonically as $g$ is increased. Therefore, in the case of two non-identical qubits and an Ohmic environment, the system will lie within the QZE regime, for any value of $g$.

We believe that the significance of our results rests with the synthesis of exceptional Points in the presence of non-Markovian dissipation. Although both aspects represent problems of extensive current research activity, their combined effect in the same quantum system has hardly been explored. Yet, the dynamics of open quantum systems associated with the presence of EPs or/and the regions of the onset of the QZE beyond Markovianity is of great significance in a plethora of realistic situations, many of practical interest.  Our method can account for any form of the boundary environment's spectral density and can easily be generalized to open quantum systems consisting of qubits and environments interacting in more complex arrangements. The drastic differences in the effective decay rate of the first qubit as a function of $g$, between a boundary environment of Lorentzian spectral density from that of Ohmic, raises the profound question: What are the necessary conditions that an arbitrary spectral density should satisfy, in order to entail the existence of EPs for certain regions of parameters? Whether these conditions are related to symmetries of the spectral density profile or other features remains to be examined in future work, with possibly quite impactful implications. Be that as it may, the contrast between the effect of a Lorentzian and an Ohmic spectral density on the EPs in our system, illustrates the resistance of non-Markovian distribution to general classifications. This seems to be a fundamental issue that has come up in ongoing related work of ours, presently in the final stage.

\section*{Acknowledgments}
GM would like to acknowledge the Hellenic Foundation for Research and Innovation (HFRI) for financially supporting this work under the 3rd Call for HFRI PhD Fellowships (Fellowship Number: 5525). We are also grateful to T. Ilias for useful discussions concerning this work.

\begin{figure}[H] 
	\centering
	\includegraphics[width=5cm]{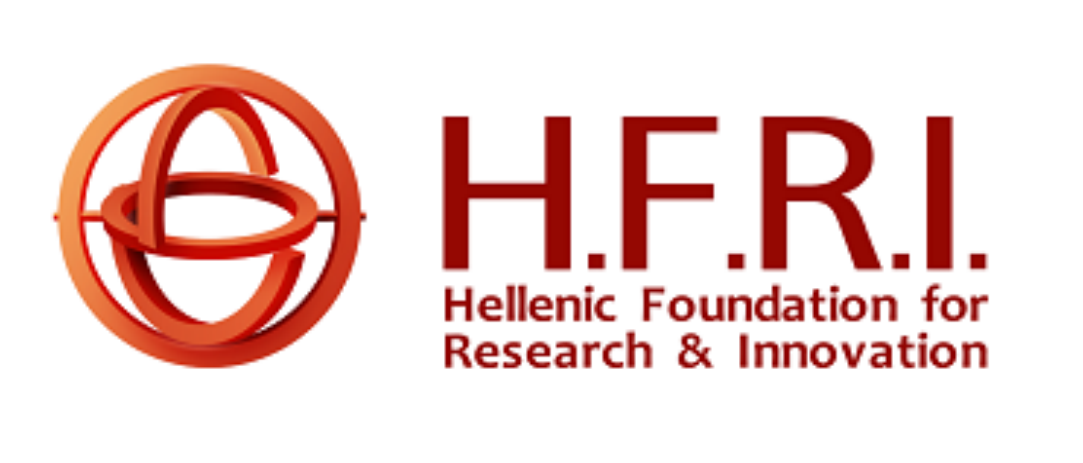}
\end{figure}

\clearpage

\appendix
\renewcommand{\thesection}{\Alph{section}}
\numberwithin{equation}{section}

\end{document}